\documentclass[3p]{elsarticle}
\journal{Radiology AI}

\usepackage[utf8]{inputenc} 
\usepackage[T1]{fontenc}    
\usepackage{textcomp}
\usepackage{helvet}
\usepackage{url}            
\usepackage{booktabs}       
\usepackage{multirow}
\usepackage{amsfonts}       
\usepackage{amsmath}        %
\usepackage{amssymb}        %
\usepackage{amsthm}         %
\usepackage{nicefrac}       
\usepackage{microtype}      
\usepackage{xcolor}         
\usepackage{lipsum}  
\usepackage{graphicx}
\usepackage{epsfig}
\usepackage{pgfplots}
\pgfplotsset{compat=newest}
\usepackage[labelfont=bf]{caption}
\usepackage[caption=false]{subfig}
\usepackage{float}
\usepackage{lineno}

\begin{document}

\begin{frontmatter}

\title{Segment-and-Classify: ROI-Guided Generalizable \\Contrast Phase Classification in CT Using XGBoost}

\author[1]{Benjamin Hou\corref{corrauth}}
\author[3]{Tejas Sudharshan Mathai}
\author[3]{Pritam Mukherjee}
\author[3]{Xinya Wang}
\author[3]{\\Ronald M. Summers}
\author[2]{Zhiyong Lu}

\address[1]{Division of Intramural Research, National Library of Medicine -}
\address[2]{National Center for Biotechnology Information, National Library of Medicine -}
\address[3]{Imaging Biomarkers and Computer Aided Diagnosis Lab, Clinical Center -
\\- National Institutes of Health, Bethesda, MD, USA.}


\begin{abstract} 



\noindent\textbf{Purpose:} To automate contrast phase classification in CT using organ-specific features extracted from a widely used segmentation tool with a lightweight decision tree classifier.

\noindent\textbf{Materials and Methods:} This retrospective study utilized three public CT datasets from separate institutions. The phase prediction model was trained on the WAW-TACE (median age: 66 [60,73]; 185 males) dataset, and externally validated on the VinDr-Multiphase (146 males; 63 females; 56 unk) and C4KC-KiTS (median age: 61 [50.68; 123 males) datasets. Contrast phase classification was performed using organ-specific features extracted by TotalSegmentator, followed by prediction using a gradient-boosted decision tree classifier.

\noindent\textbf{Results:} On the VinDr-Multiphase dataset, the phase prediction model achieved the highest or comparable AUCs across all phases (>0.937), with superior F1-scores in the non-contrast (0.994), arterial (0.937), and delayed (0.718) phases. Statistical testing indicated significant performance differences only in the arterial and delayed phases (p<0.05). On the C4KC-KiTS dataset, the phase prediction model achieved the highest AUCs across all phases (>0.991), with superior F1-scores in arterial/venous (0.968) and delayed (0.935) phases. Statistical testing confirmed significant improvements over all baseline models in these two phases (p<0.05). Performance in the non-contrast class, however, was comparable across all models, with no statistically significant differences observed (p>0.05).

\noindent\textbf{Conclusion:} The lightweight model demonstrated strong performance relative to all baseline models, and exhibited robust generalizability across datasets from different institutions.

\end{abstract}


\end{frontmatter}


\newpage

\noindent\textbf{Summary Statement:}

\noindent By utilizing contrast-relevant organ features, extracted using TotalSegmentator, with a lightweight decision tree classifier, the resulting contrast phase prediction model achieves performance comparable to 3D CNNs and demonstrates robustness to domain shifts.

\vspace{\baselineskip}

\noindent\textbf{Key Points:}

\begin{enumerate}
    \item This study evaluates a lightweight contrast phase classification model across three public datasets, training on one (WAW-TACE) and validating performance on two independent external datasets from different institutions (VinDr-Multiphase and C4KC-KiTS).
    \item The lightweight model demonstrates significantly superior performance in several phase classes, otherwise comparable performance with no statistically significant differences.
    \item All models performed comparably on the non-contrast phase, with no statistically significant differences.
\end{enumerate}

\vspace{\baselineskip}

\noindent\textbf{Abbreviations:}

\noindent $-\hspace{0.1em}$ GBDT - Gradient Boosted Decision Tree \\
\noindent $-\hspace{0.1em}$ HCC - Hepatocellular Carcinoma \\
\noindent $-\hspace{0.1em}$ IV - Intravenous \\
\noindent $-\hspace{0.1em}$ TS - TotalSegmentator \\




\vspace{\baselineskip}

\noindent\textbf{Acknowledgments:}

\noindent This research was supported by the Intramural Research Program of the National Institutes of Health (NIH); National Library of Medicine (NLM) and Clinical Center (CC). This work utilized the computational resources of the NIH high-performance computing Biowulf cluster (\url{http://hpc.nih.gov}).




\newpage

\section{Introduction}
\label{sec_intro}

CT imaging with intravenous (IV) contrast is a routine exam that is widely performed for assessing organ function (e.g., liver and kidney function), identifying the cause of traumatic injuries \cite{Rudie2024}, and diagnosing various diseases (e.g., lymphoma, liver cancer) \cite{rogers2023intravenous,Radetic595}. The CT exam captures anatomical changes and physiological processes through the acquisition of multiple CT series \cite{boraschi2020dynamic,baron1994understanding}. Before administration of contrast, a non-contrast CT series is typically acquired and is useful for assessing atherosclerotic plaque and kidney stones. Post-administration of IV contrast, the contrast material circulates through the body and highlights various anatomical structures that enhance differently. During the arterial phase, the contrast highlights the arterial system, providing detailed images of vasculature and is useful for visualizing any enhancing lesions that have an arterial supply (e.g., hepatocellular carcinoma or HCC). Next, the portal-venous phase is useful to determine attenuation differences between a lesion and the surrounding tissue \cite{boraschi2020dynamic}. Finally, in the delayed or excretory phase, the contrast is processed through the kidneys, allowing for visualization of the urinary tract and assessment of renal function.

Certain CT phases are often read in conjunction by a radiologist since they provide complementary information related to disease status. For example, both arterial and portal-venous phases are commonly used for diagnosing HCC \cite{nayantara2020computer,militzer2010automatic}. At the time of CT acquisition, details regarding the phase being acquired are entered into the DICOM header, such as ``Body Part Examined'', ``Procedure Step Description'', ``Series Description'', and ``Protocol Name''. Text-based rules are set on the DICOM header tags that dictate the arrangement of different CT series on the PACS viewer according to the radiologists' viewing preference, otherwise known as the hanging protocol. However, $\sim$16\% of DICOM headers have inaccurate information \cite{gueld2002quality,kim2025automated,harvey2019standardised} arising from heterogeneous and inconsistent data entry, which hinders the use of DICOM tags for automatic series arrangement. Typically, radiologists correct the series arrangement by dragging and dropping the series to their preferred order. But, this process takes $\sim$1-2 minutes depending on the study type, and during a busy clinical day where $\geq$40 studies are to be read, this corresponds to $\sim$40 minutes (in an 8-hour day) spent to correct the series arrangement \cite{kim2025automated}.

Data orchestration is crucial for deploying AI algorithms that depend on specific CT series or studies. Since textual guidelines in DICOM headers are sometimes not adhered to, using the wrong AI algorithm for a given task can lead to inaccurate results. With the rising volume of CT exams in the U.S. \cite{mahesh2022patient}, automation could help alleviate radiologist burnout \cite{thakore2024best}. However, most AI models rely on curated datasets \cite{kim2025automated,harvey2019standardised}, which lacks phase-specific annotations despite dense labeling of 117 structures. Identifying series types manually for small datasets (<2000 studies) is manageable, but this becomes impractical as CT volumes continue to grow.

This study addresses the challenges of contrast phase classification by combining a well-established organ segmentation model, TotalSegmentator (TS) \cite{Wasserthal2024}, with a gradient-boosted decision tree (GBDT) classifier, XGBoost \cite{chen2016xgboost}, to create an efficient and robust pipeline. TS is used to segment contrast-relevant organs, from which regional intensity statistics are extracted and used as features for phase prediction. This approach offers notable advantages over conventional 3D convolutional neural networks (CNNs), as it eliminates the need to train complex deep learning models. Experimental results demonstrate that the classifier not only outperforms traditional deep learning-based methods, but also exhibits strong generalizability under distribution shifts, as validated on two external datasets.

\section{Materials and Methods}
\label{sec_methods}

\noindent
This retrospective study employed three publicly available datasets; WAW-TACE \cite{bartnik2024waw}, VinDr-Multiphase \cite{dao2022phase}, and C4KC-KiTS \cite{heller2019kits}. Table \ref{tab:data_characteristics} and \ref{tab:phase_dist} show the dataset characteristics. All datasets complied with the Health Insurance Portability and Accountability Act (HIPAA), and received approval for research use from the Institutional Review Boards (IRB). The requirement for signed informed consent was waived. 

\subsection{Patient Sample}

The WAW-TACE dataset \cite{bartnik2024waw}, developed for research in HCC research, comprises 233 patients (median age: 66 [60,73]; 185 males) and was collected at the Medical University of Warsaw between 2016 and 2021. It contains a total of 854 CT scans annotated with four contrast-phase labels: Non-contrast, Arterial, Venous, and Delayed. This dataset is used as the training set for the phase classification models.

The VinDr-Multiphase dataset \cite{dao2022phase}, developed to support research in abdominal CT phase recognition, consists of 265 studies from 265 patients (mean age unspecified; 146 male, 63 female, 56 unknown), and is collected from the PACS databases of two Vietnamese hospitals between 2015 and 2020. It contains a total of 1,188 CT scans annotated with four contrast-phase labels: Non-contrast, Arterial, Venous, and Others. This dataset is used as an external test set for evaluating the phase classification models. 

The C4KC-KiTS dataset \cite{heller2019kits}, originally released for the KiTS19 challenge, comprises 210 contrast-enhanced abdominal CT scans acquired in the corticomedullary phase. In total, 408 scans were collected during routine clinical care of patients undergoing partial or radical nephrectomy at the University of Minnesota Medical Center. The dataset includes three contrast-phase labels: Non-contrast, Arterial, and Late. C4KC-KiTS is used as a secondary external test set to evaluate the performance of phase classification models.

\subsection{TotalSegmentator}

TotalSegmentator \cite{Wasserthal2024}, built upon the nnU-Net deep learning framework \cite{isensee2021nnu}, is an open-source tool for automatic multi-organ segmentation of CT scans. A common approach in literature for contrast phase prediction involves first segmenting the organs of interest, i.e., the region of interest (ROI), followed by applying a simple classification or regression model based on intensity-derived features \cite{liu2024utilizing,reis2024automated,salimi2024fully,baldini2023addressing}. TS includes an auxiliary script (totalseg\_get\_phase.py) that follows this strategy by segmenting 16 key organs that best capture the anatomical and physiological context relevant to phase classification. The intensity-derived features are then used as input to an XGBoost \cite{chen2016xgboost} regressor to predict the post-injection time (\texttt{pi\_time}), which is subsequently mapped to discrete phase classes based on predefined timing boundaries. This baseline comparative method is thereon referred to as ts\_phase.

\begin{figure}[H]
    \centering
    \includegraphics[width=0.65\linewidth]{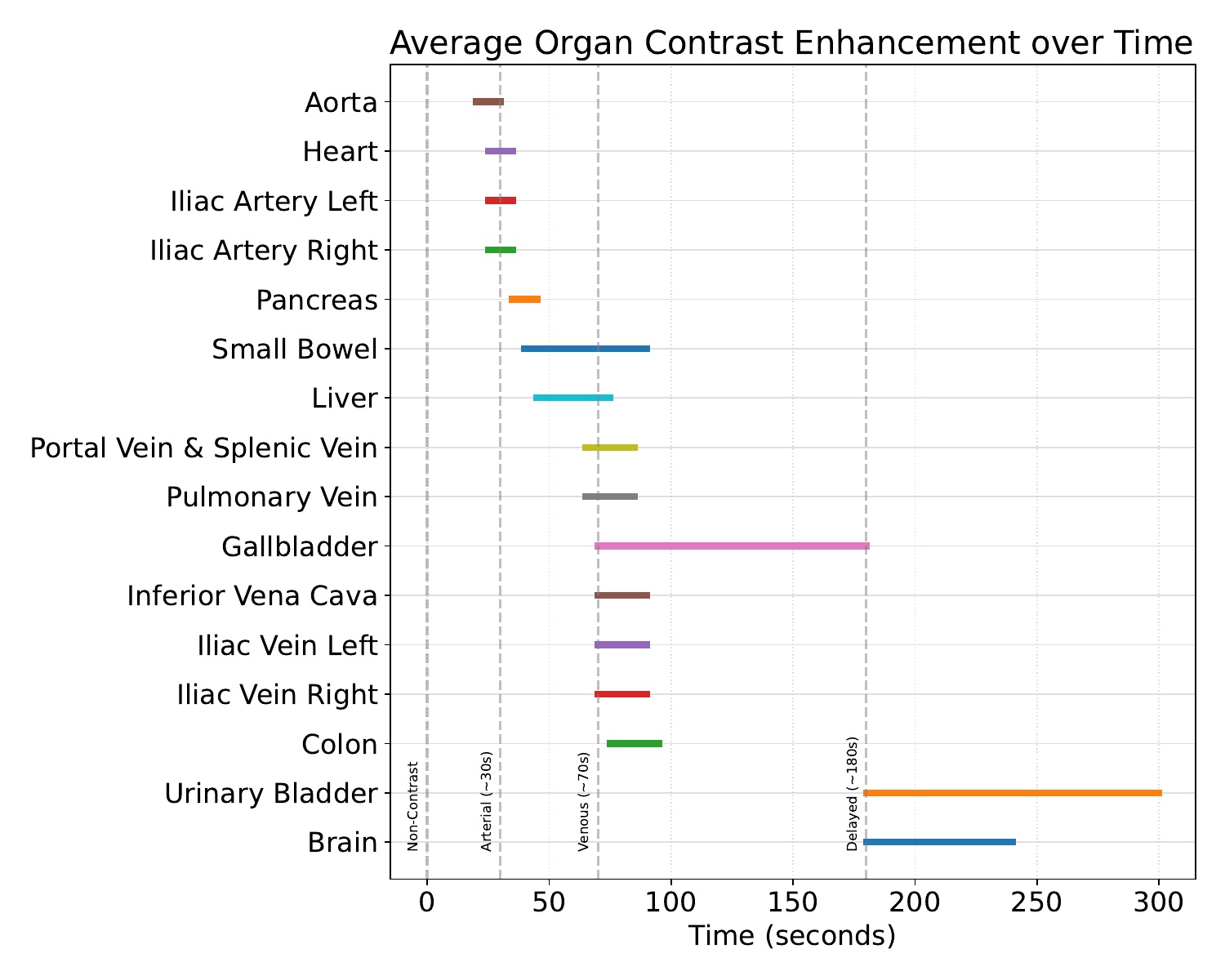}
    \caption{Contrast enhancement timing of selected organs from the TotalSegmentator list. Organs are sorted by onset of enhancement. Horizontal lines denote time post-contrast injection ($t=0$) during which each organ is best visualized. Dashed vertical lines indicate standard CT phases timings: non-contrast, arterial ($\sim$30s), venous ($\sim$70s), and delayed ($\sim$180s).}
    \label{fig:phase-timing-graph}
\end{figure}

The segmented organs includes; liver, pancreas, urinary bladder, gallbladder, heart, aorta, inferior vena cava, portal vein and splenic vein, left and right iliac veins, left and right iliac arteries, pulmonary vein, brain, colon, and small bowel. Figure \ref{fig:phase-timing-graph} illustrates the timing of optimal contrast visualization for each organ relative to standard CT phases. These regions serve as targeted inputs for subsequent analysis, helping to reduce irrelevant background information and focus the model on clinically meaningful structures. As this study focuses on abdominal imaging, head and neck structures such as the internal carotid arteries (left and right) and internal jugular veins (left and right) are excluded from analysis. 

\subsection{XGBoost Classifier}

Unlike ts\_phase, this study uses an XGBoost \cite{chen2016xgboost} classifier to predict the contrast phase label. After segmenting key organs, the median intensity within each ROI is extracted, resulting in a 16-dimensional feature vector where each each dimension corresponding to one of the segmented organs. Also in contrast to TS, if an organ is absent or not fully captured in the scan, its corresponding feature is set to NaN rather than zero. This leverages XGBoost's native ability to handle missing values by learning optimal split directions for missing features during training. The phase classification model is publicly available at \url{https://github.com/farrell236/CTPhase-XGBoost}.

\subsection{Baseline 3D CNNs}

Three standard 3D convolutional neural networks (CNNs) were used as baseline models: ResNet3D (r3d\_18) \cite{du2021revisiting}, Mixed Convolution Network (mc3\_18) \cite{tan2019mixconv}, and R(2+1)D (r2plus1d\_18) \cite{tran2018closer}, each configured with 18 layers by default. The input CT volumes were resampled to an isotropic voxel spacing of 1.6 $\times$ 1.6 $\times$ 1.6 mm and center-cropped to 240 $\times$ 240 $\times$ 240 voxels. Intensity values were windowed using a center of –400 and a width of 1400. 

\subsection{Training and Evaluation}

Both the XGBoost and 3D CNN models were trained using five-fold cross-validation on the WAW-TACE dataset. To ensure balanced class distributions and no data leakage, `StratifiedGroupKFold' from the \texttt{scikit-learn} library (v1.6.1) \cite{pedregosa2011scikit} was used. This method stratifies by phase label and groups by patient ID, ensuring that all scans from a given patient are contained within a single fold. Five splits were used (\texttt{n\_splits=5}), with shuffling enabled and a fixed random seed (\texttt{random\_state=42}) for reproducibility. The weights of the best-performing model from each fold were saved.

The XGBoost classifier was trained with a learning rate of 0.05, a maximum tree depth of 4, and 200 estimators. The evaluation metric was set to \texttt{mlogloss} to reflect the four-class phase classification task. All available CPU cores were utilized (\texttt{n\_jobs=-1}), and training completed in under one minute.

The 3D CNN models were trained using the Adam optimizer with a learning rate of 1e–5 for 20 epochs. ResNet3D and Mixed Convolution were trained with a batch size of 4, while R(2+1)D used a reduced batch size of 2 due to memory constraints. The average training time per epoch was approximately 3 minutes for ResNet3D and Mixed Convolution, and 11 minutes for R(2+1)D. All training was conducted on an NVIDIA A100 80GB GPU using PyTorch (v2.4.1) and Python (v3.9.15).

The trained models were then evaluated on the VinDr-Multiphase and C4KC-KiTS datasets. For each fold, the best-performing model was used to generate logits, which were then averaged across folds to produce a soft ensemble prediction. The final predicted class for each sample was determined by taking the \texttt{argmax} over the averaged logits. Evaluation metrics included the area under the ROC curve (AUC), sensitivity, specificity, positive predictive value (PPV), F1-score, and overall accuracy.

\subsection{Statistical Analysis}


To assess whether differences in model performance were statistically significant, McNemar’s test was applied to compare classification outcomes on a per-class basis across all model pairs. This approach helps control for class imbalance across classes, as it isolates performance comparisons within each class rather than averaging across an imbalanced distribution. However, if there is substantial imbalance within a class (e.g., very few samples for a particular phase), the test may lack statistical power or yield unreliable results. The analysis was performed using the `statsmodels' library (v0.14.4), and a p-value less than 0.05 was considered statistically significant. Comparisons with no discordant predictions (i.e., where both models always agreed) were marked as NaN, as McNemar’s test could not be performed in these cases.

\section{Results}
\label{sec_results}

\subsection{Cross-Dataset Label Adjustment}

Since the model was trained on the WAW-TACE dataset, it is capable of classifying four contrast phases: non-contrast, arterial, venous, and delayed. However, the VinDr-Multiphase dataset does not include an explicit delayed phase. Instead, it features a category labeled other, which, according to the dataset description, ``refers to all scans that cannot be correctly classified as either of [the] 3 phases ... [and] may include scans of the delay phase or scans that belong to a transitional state between 2 phases'' \cite{dao2022phase}. As a result, a performance drop is expected for the delayed class during evaluation.

The C4KC-KiTS dataset includes only three phase labels: non-contrast, arterial, and delayed \cite{heller2019kits}. To ensure compatibility with the phase classifier model, both arterial and venous predictions were relabeled as a single arterial class during evaluation. Correspondingly, the ground truth arterial labels in C4KC-KiTS were interpreted as encompassing the entire post-contrast arterial-to-venous range. It is also worth noting that some scans in the dataset may have been misclassified; specifically, certain scans labeled as arterial were identified by experts as more consistent with venous phase imaging \cite{baldini2023addressing}.

\subsection{Results on VinDr-Multiphase}

Table \ref{tab:results_vindr} presents the phase classification results on the VinDr-Multiphase, Figure \ref{fig:roc_cm_vindr} and \ref{fig:cm_ts_phase}a shows the confusion matrix and ROC for each model. XGBoost achieved generally high performance across all phases. For the non-contrast phase, it reached near perfect scores for all metrics. However, all other models also demonstrated strong performance, and accuracy differences were mostly not statistically significant (p=0.47 for both r3d\_18 and r2plus1d\_18; p=0.11 for ts\_phase). Notably, mc3\_18 showed identical accuracy to XGBoost with no discordant predictions, and therefore no p-value was computed.

In the arterial phase, XGBoost maintained strong performance with high scores for AUC (0.97), specificity (0.99), PPV (0.99), and F1-score (0.93), although sensitivity and accuracy were comparatively lower at 0.88 for both. Model performance varied more substantially in this phase, but XGBoost still achieved the highest AUC, accuracy, and F1-score among all models, with statistically significant differences in accuracy observed for all comparisons (p<0.05).

In the venous phase, XGBoost achieved the highest performance in AUC (0.97), sensitivity (0.93), and accuracy (0.93), but was outperformed by mc3\_18 in specificity (0.96), PPV (0.93), and F1-score (0.93). The other models also demonstrated strong performance, with no statistically significant differences in accuracy observed across any comparisons (p>0.05).

For the delayed phase, all models exhibited reduced performance with greater variability, attributable to the heterogeneous ``other'' class in the VinDr-Multiphase dataset. In the delayed phase, XGBoost showed a higher specificity (0.96) than sensitivity (0.78), whereas the 3D models demonstrated the opposite trend, with r3d\_18, mc3\_18, and r2plus1d\_18 reaching sensitivities of 0.91, 0.86, and 0.92, but lower specificities of 0.90, 0.93, and 0.87, respectively. This pattern reflects a more conservative classification behavior by XGBoost, which led to a higher PPV (0.66) and F1-score (0.71), but a lower overall accuracy (0.78) compared to the 3D models (r3d\_18: 0.91, mc3\_18: 0.86, r2plus1d\_18: 0.92), with the differences in accuracy being statistically significant only for r3d\_18 and r2plus1d\_18 (p<0.05). ts\_phase model completely failed to identify any delayed-phase cases correctly; the confusion matrix (Figure \ref{fig:cm_ts_phase}a) indicates that all delayed-phase scans were misclassified as either arterial or venous.

\subsection{Results on C4KC-KiTS}

Table \ref{tab:results_c4kckits} presents the phase classification results on the VinDr-Multiphase, Figure \ref{fig:roc_cm_c4kckits} and \ref{fig:cm_ts_phase}b shows the confusion matrix and ROC for each model. Among all models, XGBoost demonstrated the strongest overall performance, achieving values above 0.95 across all metrics, except for PPV (0.915) and F1-score (0.935) in the delayed phase. For the non-contrast phase, XGBoost achieved near-perfect scores across all metrics, including an AUC of 0.99, sensitivity of 0.98, specificity of 0.99, and accuracy of 0.98. All other models also demonstrated strong performance, and accuracy differences were mostly not statistically significant (p=1.00 for both mc3\_18 and ts\_phase). Notably, r3d\_18 and r2plus1d\_18 showed identical accuracy to XGBoost with no discordant predictions, and therefore no p-value was computed.

For the arterial/venous phase, XGBoost again achieved the highest scores across all metrics, with an AUC of 0.99, sensitivity of 0.96, specificity of 0.97, PPV of 0.98, F1-score of 0.97, and accuracy of 0.96. Performance among the other models was notably lower across all metrics, and all accuracy differences compared to XGBoost were statistically significant (p<0.05 for all models). The confusion matrices for the 3D models and ts\_phase revealed frequent misclassification of arterial/venous scans as delayed phase, contributing to reduced AUC and F1-scores.

For the delayed phase, XGBoost consistently demonstrated robust performance, achieving an AUC of 0.99, sensitivity of 0.95, specificity of 0.97, and an F1-score of 0.93. In contrast, the 3D models and ts\_phase showed markedly lower performance. r3d\_18, mc3\_18, and r2plus1d\_18 achieved F1-scores of 0.68, 0.47, and 0.50, respectively, while ts\_phase performed the worst with an F1-score of 0.18. These values were substantially lower than their corresponding F1-scores in the arterial/venous phase, indicating a consistent drop in performance when predicting the delayed phase. Similar to its performance on the VinDr-Multiphase dataset, ts\_phase failed to accurately predict the delayed class in the C4KC-KiTS dataset. The majority of delayed-phase scans were misclassified as arterial/venous, as reflected in the confusion matrix.

\section{Discussion}
\label{sec_discussion}

This study addressed the challenge of contrast phase classification in CT by leveraging TS to segment and quantify organ-specific intensity features relevant to contrast phase, combined with a lightweight GBDT classifier, XGBoost, for phase prediction. The model demonstrated strong performance on the external VinDr-Multiphase dataset, achieving high AUCs and F1-scores, and significantly outperformed both 3D CNN models and the ts\_phase classifier across the majority of phase classes. Its generalizability was further confirmed on a second external dataset, C4KC-KiTS, where it consistently maintained high performance and outperformed all comparative models.

Related works address this problem using three primary methods: (1) resampling entire 3D volumes to smaller dimensions for training 3D networks, (2) randomly sampling 2D slices and using majority voting to determine the phase class, or (3) employing segmentation models to delineate specific contrast-related organs. The first method is straightforward and often effective \cite{ZhouHYCXL019, TangLXTCGHGBSAH20}, but training 3D models is computationally expensive, and downsampling sacrifices native CT resolution. The second method, training 2D networks with individual slices, is more common in the literature \cite{rocha2022contrast, muhamedrahimov2022using, anand2023automated, dao2022phase} but has limitations since phase labels apply to entire volumes, not slices. Assigning volume-level labels to slices risks noisy classification, especially when acquisition phases overlap. For instance, distinguishing portal venous from delayed phases using only a chest slice is infeasible without specific imaging features. To mitigate this, some studies propose localizing relevant slices before classification or sampling multiple slices for majority voting. The third approach relies on segmentation models to delineate contrast-related organs \cite{salimi2024fully,LIU2024102458} or extract radiomics features \cite{reis2023automatic}, followed by classification. 

\begin{figure}[H]
    \centering
    \includegraphics[width=0.25\linewidth]{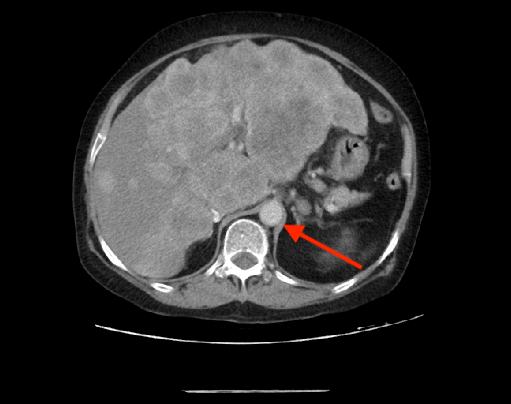}
    \includegraphics[width=0.25\linewidth]{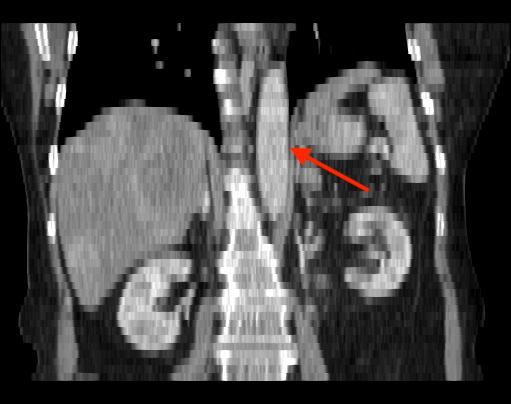}
    \includegraphics[width=0.25\linewidth]{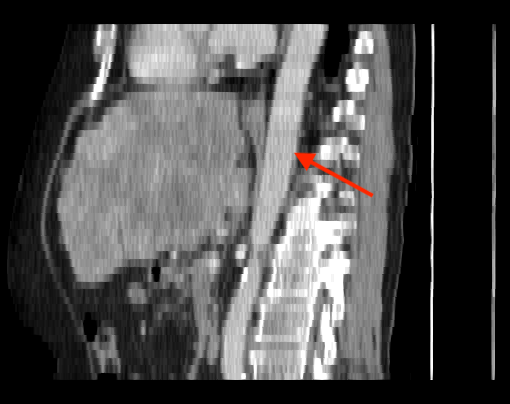}
    \caption{Arterial-phase scan of Patient \#1033 from the VinDr-Multiphase dataset. XGBoost correctly classified the scan as arterial, while all 3D models predicted delayed, and ts\_phase misclassified it as venous. This case is particularly challenging, as the aorta shows only subtle enhancement—possibly corresponding to a late arterial phase—a feature that was overlooked by the 3D models.}
    \label{fig:vindr1033}
\end{figure}

The advantage of ROI-based segmentation and feature extraction is that it directs the model's attention to clinically relevant regions, reducing the influence of irrelevant surrounding anatomy. In contrast, 3D models process the entire volume, where non-contributing structures can introduce residual noise that may hinder phase prediction performance, e.g. Figure \ref{fig:vindr1033}. To ensure consistency across inputs, 3D volumes are also resampled to isotropic spacing and center-cropped to a fixed size. While it is possible to train models without resampling, this introduces additional variability of pixel spacings. 
These trade-offs must be carefully considered during model design.

ts\_phase demonstrated the utility of segmentation-based phase prediction by regressing \texttt{pi\_time}. However, the model performs poorly in predicting the delayed phase, a limitation acknowledged by the authors and reflected in our evaluation, where delayed scans were frequently misclassified as venous. In this study, we address this limitation by training models directly on phase class labels rather than regressed time values. Additionally, our approach enables the estimation of a pseudo \texttt{pi\_time} by computing the dot product of class logits with predefined phase boundary times. Currently, no dataset in the literature includes ground-truth \texttt{pi\_time} annotations. Since DICOM metadata is often inconsistent and unreliable, careful manual curation would be required to generate such a dataset.

\begin{figure}[H]
    \centering
    \includegraphics[width=0.25\linewidth]{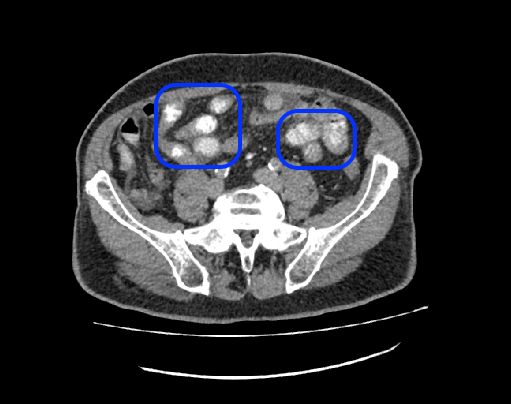}
    \includegraphics[width=0.25\linewidth]{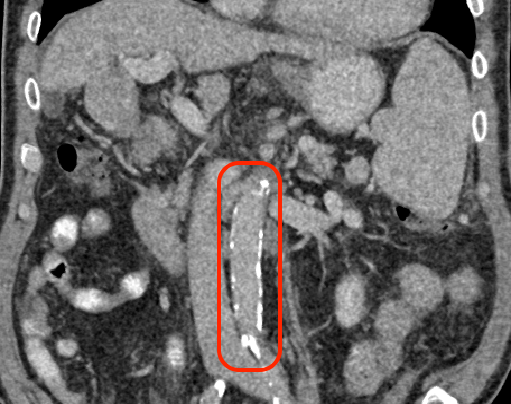}
    \includegraphics[width=0.25\linewidth]{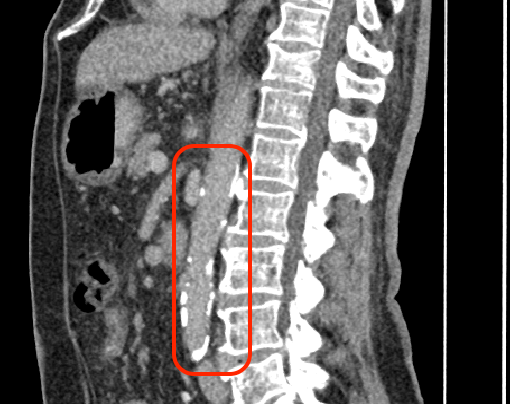}
    \caption{Delayed-phase scan of Patient \#349 from the WAW-TACE dataset. The image is limited to the abdomen, with no visualization of the urinary bladder; however, residual contrast is visible in the small bowel (blue box), and the inferior vena cava (IVC) (red box).}
    \label{fig:delayed_example}
\end{figure}

One of the central challenges in contrast phase prediction is the accurate identification of the delayed phase. This phase is most reliably recognized by assessing contrast accumulation in the urinary bladder; however, this becomes difficult when the scan does not include the pelvic region, resulting in the bladder being absent, e.g., in Figure \ref{fig:delayed_example}. In such cases, distinguishing between non-contrast and delayed phases is particularly challenging—a limitation that affects not only our model but all phase classification approaches. Therefore, differentiation may depend on more subtle imaging features, such as faint contrast retention along the gastrointestinal tract or delayed enhancement in solid organs.

Table \ref{tab:phase_dist} illustrates the distribution of phase classes across different body regions. In the VinDr-Multiphase dataset, the ``other'' category comprises only 102 out of 1,188 total scans, with just 8 scans containing visible pelvic anatomy (1 PEL, 6 ABD-PEL, and 1 CH-ABD-PEL). Given that this category may also include transitional phase states, as noted by the authors, it remains unclear whether these 8 scans represent true delayed phases, leading to substantial under-representation. In contrast, the WAW-TACE dataset contains 27 delayed-phase scans (ABD-PEL) out of 854, which, while still imbalanced, offers better representation than VinDr-Multiphase and is thus a more suitable dataset for training a phase classifier.

A limitation of the current model is its inability to predict contrast phase for brain CT scans, as the WAW-TACE dataset does not include any neuroimaging data. However, this limitation is unlikely to drastically impact clinical applicability. In routine brain imaging, contrast-enhanced CT is typically performed using a single delayed post-contrast phase, primarily for evaluating tumors, infections, or vascular lesions. Unlike abdominal CT, brain imaging does not follow a standardized multiphase protocol, and distinguishing between arterial, venous, and delayed phases is not clinically relevant in most neuroimaging workflows. Therefore, a binary classification—distinguishing contrast-enhanced from non-contrast scans—is generally sufficient for brain CT.

In conclusion, our study demonstrates that XGBoost classifier can accurately predict CT contrast phases across multiple datasets. This approach offers a computationally efficient and interpretable alternative to deep learning-based models, while maintaining strong generalizability. Future work could explore extending this method to estimate pseudo \texttt{pi\_time}, should appropriately annotated temporal data become available, and further adapt the framework for applications such as real-time protocol optimization or use in other imaging modalities.

\newpage

\bibliographystyle{vancouver}
\bibliography{references}

\newpage

\begin{table}[H]
\centering
\caption{Characteristics of the datasets used in this study are presented as median and interquartile ranges. Statistics for age and sex are stratified by patient. F=female, M=male. (*Note: `Delayed' phase is referred to as `Other' in VinDr-MultiPhase and `Late' in C4KC-KiTS).}
\begin{tabular}{@{}lccc@{}}
\toprule
                        & WAW-TACE          & VinDr-MultiPhase  & C4KC-KiTS         ~~ \\
\midrule   
~~ Patients             & 233               & 265               & 210               ~~ \\
~~ Age                  & 66 [60,73]        & ---               & 61 [50,68]        ~~ \\
~~ Sex                  &                   &                   &                   ~~ \\
~~~~ M                  & 185               & 146               & 123               ~~ \\
~~~~ F                  & 48                & 63                & 87                ~~ \\
~~ Pixel Spacing (mm)   & 0.77 [0.71,0.85]  & 0.70 [0.64,0.78]  & 0.79 [0.72,0.87]  ~~ \\
~~ Slice Thickness (mm) & 2.50 [1.25,2.50]  & 1.25 [1.25,5.00]  & 3.00 [2.50,5.00]  ~~ \\
~~ No. Scans            & 854               & 1,188             & 408               ~~ \\
~~ No. Slices           & 175,222           & 392,560           & 71,108            ~~ \\
~~ Contrast             &                   &                   &                   ~~ \\
~~~~ Non-contrast       & 200               & 183               & 107               ~~ \\
~~~~ Arterial           & 230               & 491               & 210               ~~ \\
~~~~ Venous             & 231               & 412               & ---               ~~ \\
~~~~ Delayed            & 193               & 102*              & 91*               ~~ \\
\bottomrule
\end{tabular}
\label{tab:data_characteristics}
\end{table}

\begin{table}[H]
\centering
\caption{Distribution of CT scans across different anatomical coverage (CH: chest, ABD: abdomen, PEL: pelvis) and contrast phases (non-contrast, arterial, venous, delayed) for the WAW-TACE, VinDr-Multiphase, and C4KC-KiTS datasets. ``None'' denotes partial scans limited to specific organs, e.g., liver-only or kidney-only coverage. (*Note: `Delayed' phase is referred to as `Other' in VinDr-MultiPhase and `Late' in C4KC-KiTS).}
\begin{tabular}{@{}lcccccccc@{}}
\toprule
                  & CH & ABD & PEL & CH-ABD & ABD-PEL & CH-ABD-PEL & None & Total ~~ \\
\midrule
~~ WAW-TACE \\
~~~~ Non-contrast &  0 & 146 &   0 &      3 &      41 &          8 &    2 &   200 ~~ \\
~~~~ Arterial     &  1 & 182 &   1 &      7 &      30 &          7 &    2 &   230 ~~ \\
~~~~ Venous       &  1 &  58 &   2 &      1 &     141 &         23 &    5 &   231 ~~ \\
~~~~ Delayed      &  1 & 164 &   0 &      0 &      27 &          0 &    1 &   193 ~~ \\
~~~~ Total        &  3 & 550 &   3 &     11 &     239 &         38 &   10 &   854 ~~ \\
\midrule
~~ Vin-Dr Multiphase \\
~~~~ Non-contrast &  0 &   0 &   0 &      0 &     154 &         29 &    0 &   183 ~~ \\
~~~~ Arterial     &  0 & 315 &   0 &      2 &     172 &          0 &    2 &   491 ~~ \\
~~~~ Venous       &  0 &  20 &   0 &      0 &     351 &         40 &    1 &   412 ~~ \\
~~~~ Delayed*     &  0 &  88 &   1 &      0 &       6 &          1 &    6 &   102 ~~ \\
~~~~ Total        &  0 & 423 &   1 &      2 &     683 &         70 &    9 &  1188 ~~ \\
\midrule
~~ C4KC-KiTS \\
~~~~ Non-contrast &  0 &  55 &   1 &      0 &      51 &          0 &    0 &   107 ~~ \\
~~~~ Arterial     &  0 &  88 &   1 &      1 &     100 &         18 &    2 &   210 ~~ \\
~~~~ Delayed*     &  0 &  31 &   7 &      0 &      53 &          0 &    0 &    91 ~~ \\
~~~~ Total        &  0 & 174 &   9 &      1 &     204 &         18 &    2 &   408 ~~ \\
\bottomrule
\end{tabular}
\label{tab:phase_dist}
\end{table}

\newpage

\begin{table}[H]
\centering
\caption{Phase classification performance of models on the VinDr-Multiphase dataset: XGBoost, ResNet3D 18-layer (r3d\_18), Mixed Convolution Network 18-layer (mc3\_18), R(2+1)D 18-layer (r2plus1d\_18), and TotalSegmentator (ts\_phase). Models are evaluated using AUC, Sensitivity, Specificity, PPV, F1 Score, and Accuracy. P-values indicate the significance of accuracy differences compared to XGBoost (p<0.001 considered significant).}
\begin{tabular}{@{}lccccccc@{}}
\toprule
~~                  &   AUC & Sensitivity & Specificity &    PPV & F1-score & Accuracy & p-value ~~\\
\midrule 
~~ Non-contrast \\ 
~~~~ XGBoost        & 0.999 &       0.994 &       0.999 &  0.994 &    0.994 &    0.994 &     --- ~~\\
~~~~ r3d\_18        & 0.995 &       0.983 &       0.996 &  0.978 &    0.980 &    0.983 &   0.479 ~~\\
~~~~ mc3\_18        & 0.999 &       0.994 &       0.994 &  0.968 &    0.981 &    0.994 &     NaN ~~\\
~~~~ r2plus1d\_18   & 0.997 &       0.983 &       0.991 &  0.952 &    0.967 &    0.983 &   0.479 ~~\\
~~~~ ts\_phase      & 0.986 &       0.972 &       1.000 &  1.000 &    0.986 &    0.995 &   0.113 ~~\\
\midrule   
~~ Arterial \\   
~~~~ XGBoost        & 0.977 &       0.885 &       0.997 &  0.995 &    0.937 &    0.885 &     --- ~~\\
~~~~ r3d\_18        & 0.960 &       0.725 &       0.991 &  0.983 &    0.834 &    0.725 &  <0.001 ~~\\
~~~~ mc3\_18        & 0.973 &       0.845 &       0.977 &  0.962 &    0.900 &    0.845 &   0.011 ~~\\
~~~~ r2plus1d\_18   & 0.963 &       0.637 &       0.995 &  0.990 &    0.775 &    0.637 &  <0.001 ~~\\
~~~~ ts\_phase      & 0.877 &       0.961 &       0.794 &  0.767 &    0.853 &    0.863 &  <0.001 ~~\\
\midrule   
~~ Venous \\   
~~~~ XGBoost        & 0.974 &       0.939 &       0.919 &  0.861 &    0.898 &    0.939 &     --- ~~\\
~~~~ r3d\_18        & 0.971 &       0.934 &       0.927 &  0.873 &    0.902 &    0.934 &   0.838 ~~\\
~~~~ mc3\_18        & 0.969 &       0.924 &       0.965 &  0.933 &    0.929 &    0.924 &   0.361 ~~\\
~~~~ r2plus1d\_18   & 0.967 &       0.927 &       0.907 &  0.841 &    0.882 &    0.927 &   0.475 ~~\\
~~~~ ts\_phase      & 0.913 &       0.871 &       0.956 &  0.913 &    0.891 &    0.926 &  <0.001 ~~\\
\midrule   
~~ Delayed \\   
~~~~ XGBoost        & 0.937 &       0.780 &       0.964 &  0.666 &    0.718 &    0.780 &     --- ~~\\
~~~~ r3d\_18        & 0.945 &       0.911 &       0.900 &  0.462 &    0.613 &    0.911 &   0.003 ~~\\
~~~~ mc3\_18        & 0.953 &       0.862 &       0.932 &  0.546 &    0.669 &    0.862 &   0.061 ~~\\
~~~~ r2plus1d\_18   & 0.957 &       0.921 &       0.875 &  0.410 &    0.567 &    0.921 &   0.002 ~~\\
~~~~ ts\_phase      & 0.500 &       0.000 &       1.000 &  0.000 &    0.000 &    0.915 &  <0.001 ~~\\
\bottomrule
\end{tabular}
\label{tab:results_vindr}
\end{table}

\begin{table}[H]
\centering
\caption{Phase classification performance of models on the C4KC-KiTS dataset: XGBoost, ResNet3D 18-layer (r3d\_18), Mixed Convolution Network 18-layer (mc3\_18), R(2+1)D 18-layer (r2plus1d\_18), and TotalSegmentator (ts\_phase). Models are evaluated using AUC, Sensitivity, Specificity, PPV, F1 Score, and Accuracy. P-values indicate the significance of accuracy differences compared to XGBoost (p<0.001 considered significant).}
\begin{tabular}{@{}lccccccc@{}}
\toprule
~~                   &   AUC & Sensitivity & Specificity &   PPV & F1-score & Accuracy & p-value ~~\\
\midrule
~~ Non-contrast \\
~~~~ XGBoost         & 0.994 &       0.981 &       0.996 & 0.990 &    0.985 &    0.981 &     --- ~~\\
~~~~ r3d\_18         & 0.992 &       0.981 &       0.973 & 0.929 &    0.954 &    0.981 &     NaN ~~\\
~~~~ mc3\_18         & 0.989 &       0.971 &       0.940 & 0.852 &    0.908 &    0.971 &   1.000 ~~\\
~~~~ r2plus1d\_18    & 0.992 &       0.981 &       0.986 & 0.963 &    0.972 &    0.981 &     NaN ~~\\
~~~~ ts\_phase       & 0.984 &       0.971 &       0.996 & 0.990 &    0.981 &    0.990 &   1.000 ~~\\
\midrule
~~ Arterial/Venous \\ 
~~~~ XGBoost         & 0.994 &       0.961 &       0.974 & 0.975 &    0.968 &    0.961 &     --- ~~\\
~~~~ r3d\_18         & 0.961 &       0.876 &       0.878 & 0.884 &    0.880 &    0.876 &  <0.001 ~~\\
~~~~ mc3\_18         & 0.925 &       0.838 &       0.772 & 0.796 &    0.816 &    0.838 &  <0.001 ~~\\
~~~~ r2plus1d\_18    & 0.917 &       0.800 &       0.777 & 0.792 &    0.796 &    0.800 &  <0.001 ~~\\
~~~~ ts\_phase       & 0.620 &       0.614 &       0.626 & 0.635 &    0.624 &    0.620 &  <0.001 ~~\\
\midrule
~~Delayed \\
~~~~ XGBoost         & 0.991 &       0.956 &       0.974 & 0.915 &    0.935 &    0.956 &     --- ~~\\
~~~~ r3d\_18         & 0.926 &       0.670 &       0.917 & 0.701 &    0.685 &    0.670 &  <0.001 ~~\\
~~~~ mc3\_18         & 0.862 &       0.406 &       0.911 & 0.569 &    0.474 &    0.406 &  <0.001 ~~\\
~~~~ r2plus1d\_18    & 0.877 &       0.494 &       0.867 & 0.517 &    0.505 &    0.494 &  <0.001 ~~\\
~~~~ ts\_phase       & 0.469 &       0.197 &       0.741 & 0.180 &    0.188 &    0.620 &  <0.001 ~~\\
\bottomrule
\end{tabular}
\label{tab:results_c4kckits}
\end{table}

\newpage

\vspace{1em}

\newpage


\setcounter{figure}{0}
\setcounter{table}{0}
\renewcommand{\thefigure}{S\arabic{figure}}
\renewcommand{\thetable}{S\arabic{table}}

\section{Supplementary Materials}

\begin{table}[H]
\centering
\caption{Per-class McNemar’s test comparing model predictions on the VinDr-Multiphase dataset. The test statistic reflects the degree of disagreement in classification outcomes between model pairs, while the p-value indicates whether this disagreement is statistically significant (p<0.001). Comparisons with no discordant predictions are marked as NaN, as McNemar’s test cannot be computed in such cases.}
\begin{tabular}{@{}cc|cccccccc@{}}
\toprule
             &              
             & \multicolumn{2}{c}{Non-contrast}                            
             & \multicolumn{2}{c}{Arterial}                                
             & \multicolumn{2}{c}{Venous}                                  
             & \multicolumn{2}{c}{Delayed} ~~\\ 
\cmidrule(l){3-10}
~~ Model 1   & Model 2      
& Stat.      & p-value 
& Stat.      & p-value 
& Stat.      & p-value 
& Stat.      & p-value ~~\\
\midrule
~~ xgboost      & r3d\_18       & 0.5   & 0.479  &  64.0  & <0.001  &  0.0  &  0.838  & 8.4   &  0.003  ~~\\
~~ xgboost      & mc3\_18       & NaN   & NaN    &   6.4  &  0.011  &  0.8  &  0.361  & 3.5   &  0.061  ~~\\
~~ xgboost      & r2plus1d\_18  & 0.5   & 0.479  & 116.1  & <0.001  &  0.5  &  0.457  & 9.3   &  0.002  ~~\\
~~ xgboost      & ts\_phase     & 2.25  & 0.133  &  30.1  & <0.001  & 13.5  & <0.001  & 76.0  & <0.001  ~~\\
~~ r3d\_18      & mc3\_18       & 0.5   & 0.479  &  50.2  & <0.001  &  0.9  &  0.342  & 3.2   &  0.073  ~~\\
~~ r3d\_18      & r2plus1d\_18  & NaN   & NaN    &  18.9  & <0.001  &  0.3  &  0.579  & 0.2   &  0.617  ~~\\
~~ r3d\_18      & ts\_phase     & 0.5   & 0.479  & 104.9  & <0.001  & 13.0  & <0.001  & 89.0  & <0.001  ~~\\
~~ mc3\_18      & r2plus1d\_18  & 0.5   & 0.479  &  98.0  & <0.001  &  0.0  &  1.000  & 5.1   &  0.023  ~~\\
~~ mc3\_18      & ts\_phase     & 2.25  & 0.133  &  44.1  & <0.001  &  8.1  &  0.004  & 84.0  & <0.001  ~~\\
~~ r2plus1d\_18 & ts\_phase     & 0.5   & 0.479  & 155.0  & <0.001  &  8.4  &  0.003  & 91.0  & <0.001  ~~\\ 
\bottomrule
\end{tabular}
\end{table}

\begin{table}[H]
\centering
\caption{Per-class McNemar’s test comparing model predictions on the C4KC-KiTS dataset. The test statistic reflects the degree of disagreement in classification outcomes between model pairs, while the p-value indicates whether this disagreement is statistically significant (p<0.001). Comparisons with no discordant predictions are marked as NaN, as McNemar’s test cannot be computed in such cases.}
\begin{tabular}{@{}cc|cccccc@{}}
\toprule
             &              
             & \multicolumn{2}{c}{Non-contrast} 
             & \multicolumn{2}{c}{Arterial/Venous} 
             & \multicolumn{2}{c}{Delayed} \\ 
\cmidrule(l){3-8}
~~ Model 1   & Model 2      
& Stat.      & p-value         
& Stat.      & p-value          
& Stat.      & p-value ~~\\
\midrule
~~ xgboost      & r3d\_18      & NaN   & NaN    & 12.0  & <0.001  & 22.3  & <0.001  ~~\\
~~ xgboost      & mc3\_18      & 0.0   & 1.000  & 20.8  & <0.001  & 48.0  & <0.001  ~~\\
~~ xgboost      & r2plus1d\_18 & NaN   & NaN    & 28.6  & <0.001  & 40.0  & <0.001  ~~\\
~~ xgboost      & ts\_phase    & 0.0   & 1.000  & 64.0  & <0.001  & 60.0  & <0.001  ~~\\
~~ r3d\_18      & mc3\_18      & 0.0   & 1.000  &  4.9  &  0.026  & 22.0  & <0.001  ~~\\
~~ r3d\_18      & r2plus1d\_18 & NaN   & NaN    & 11.2  &  0.001  & 11.2  &  0.001  ~~\\
~~ r3d\_18      & ts\_phase    & 0.0   & 1.000  & 28.8  & <0.001  & 28.0  & <0.001  ~~\\
~~ mc3\_18      & r2plus1d\_18 & 0.0   & 1.000  &  4.0  &  0.043  &  4.9  &  0.026  ~~\\
~~ mc3\_18      & ts\_phase    & 0.25  & 0.617  & 19.7  & <0.001  &  7.2  &  0.007  ~~\\
~~ r2plus1d\_18 & ts\_phase    & 0.0   & 1.000  & 12.5  & <0.001  & 13.7  & <0.001  ~~\\ 
\bottomrule
\end{tabular}
\end{table}

\begin{figure}[H]
    \centering
    \begin{tabular}{cccc}
    \includegraphics[width=0.22\linewidth]{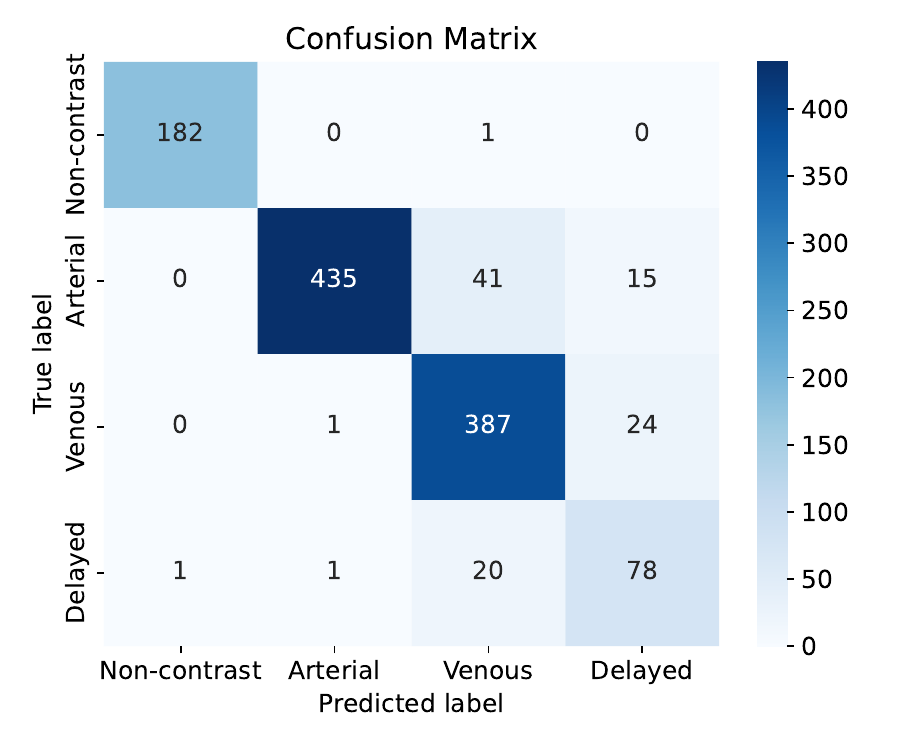} &
    \includegraphics[width=0.22\linewidth]{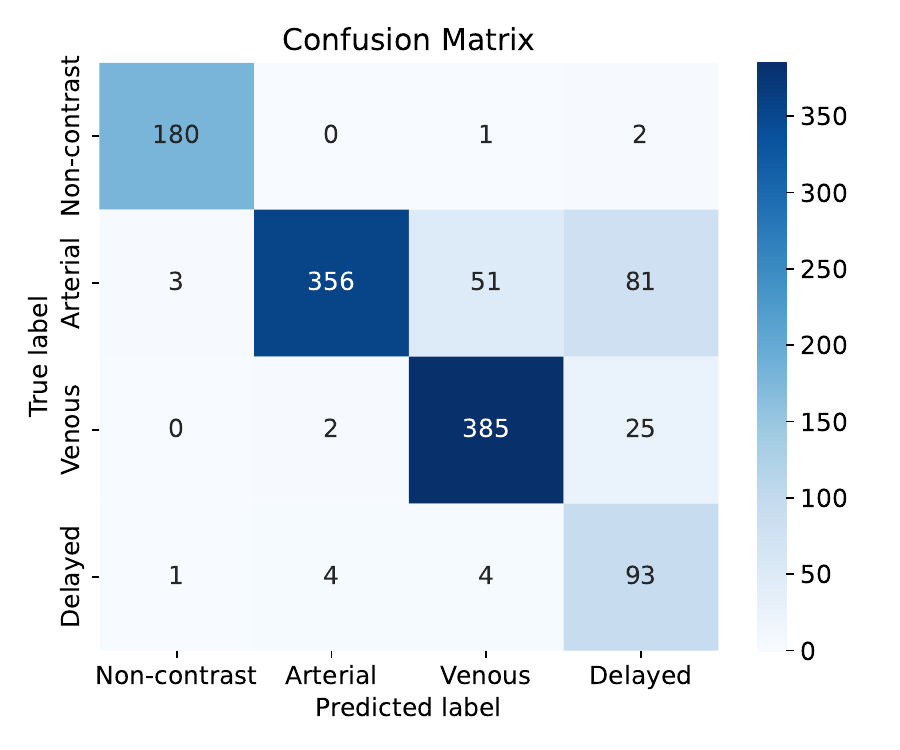} &
    \includegraphics[width=0.22\linewidth]{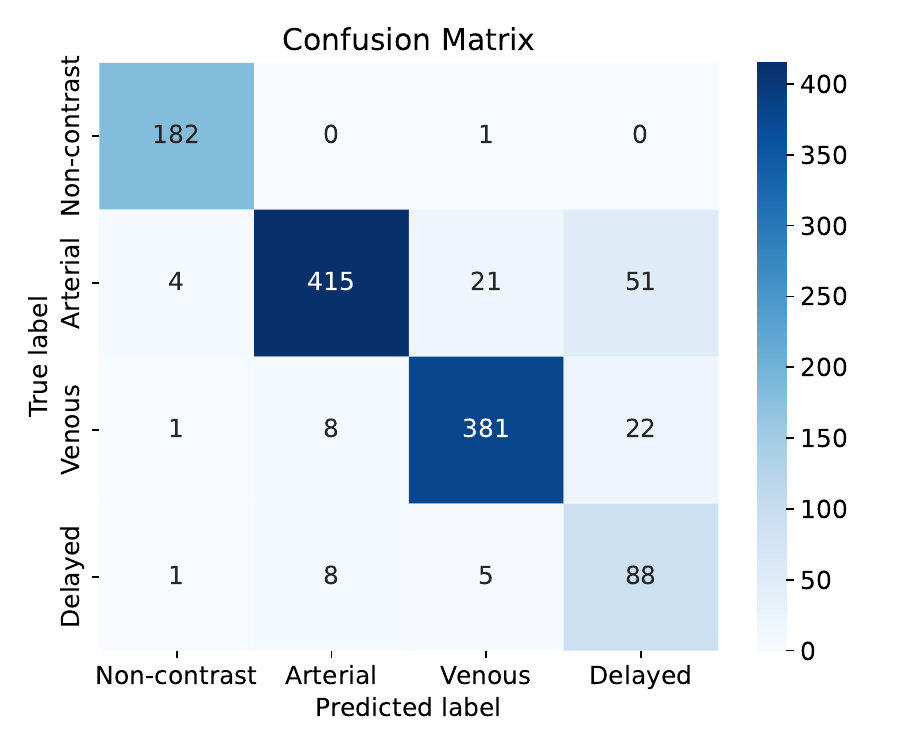} &
    \includegraphics[width=0.22\linewidth]{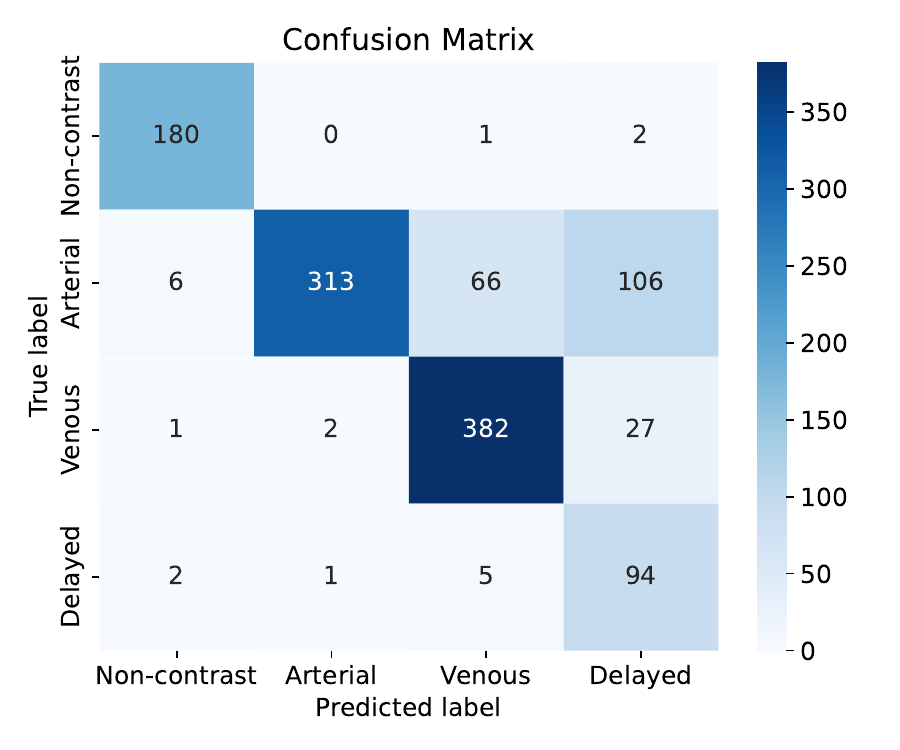} \\
    \includegraphics[width=0.22\linewidth]{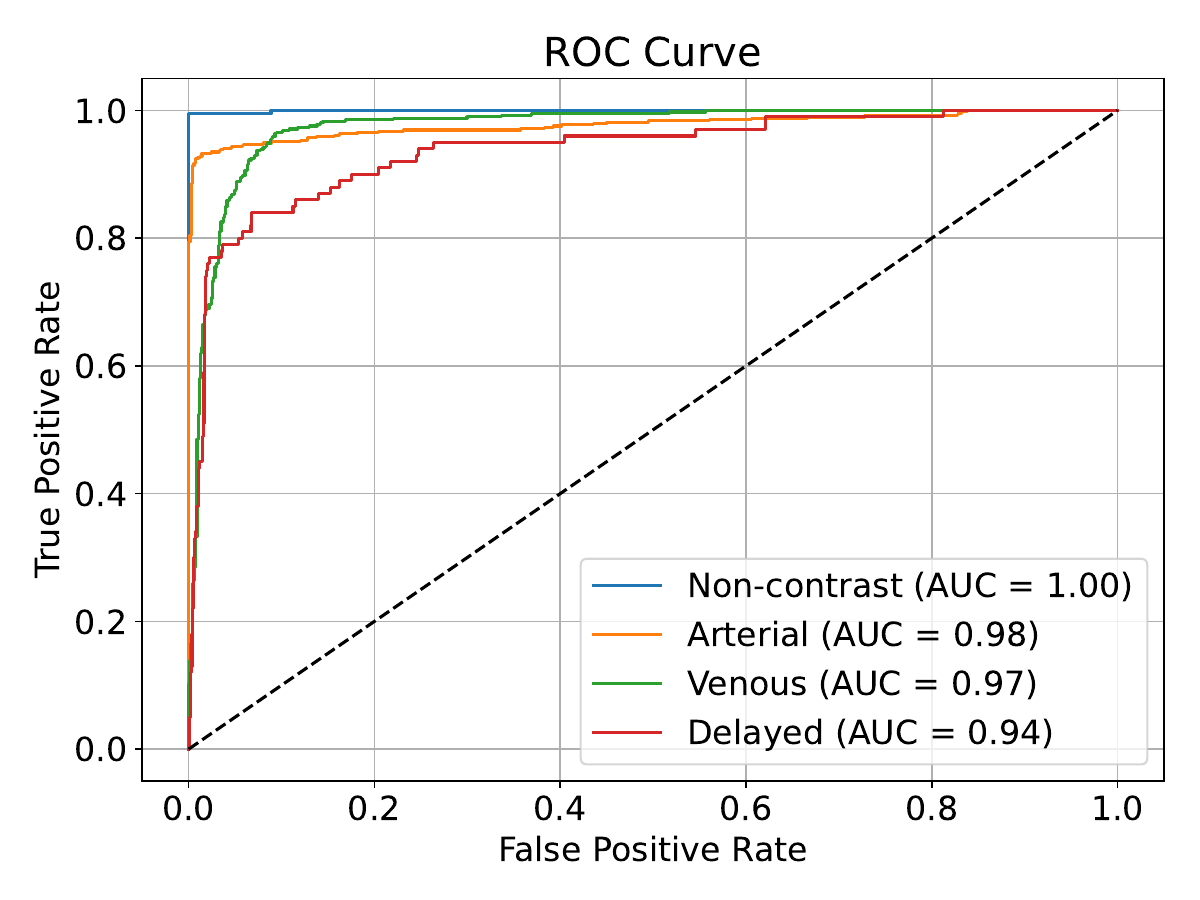} &
    \includegraphics[width=0.22\linewidth]{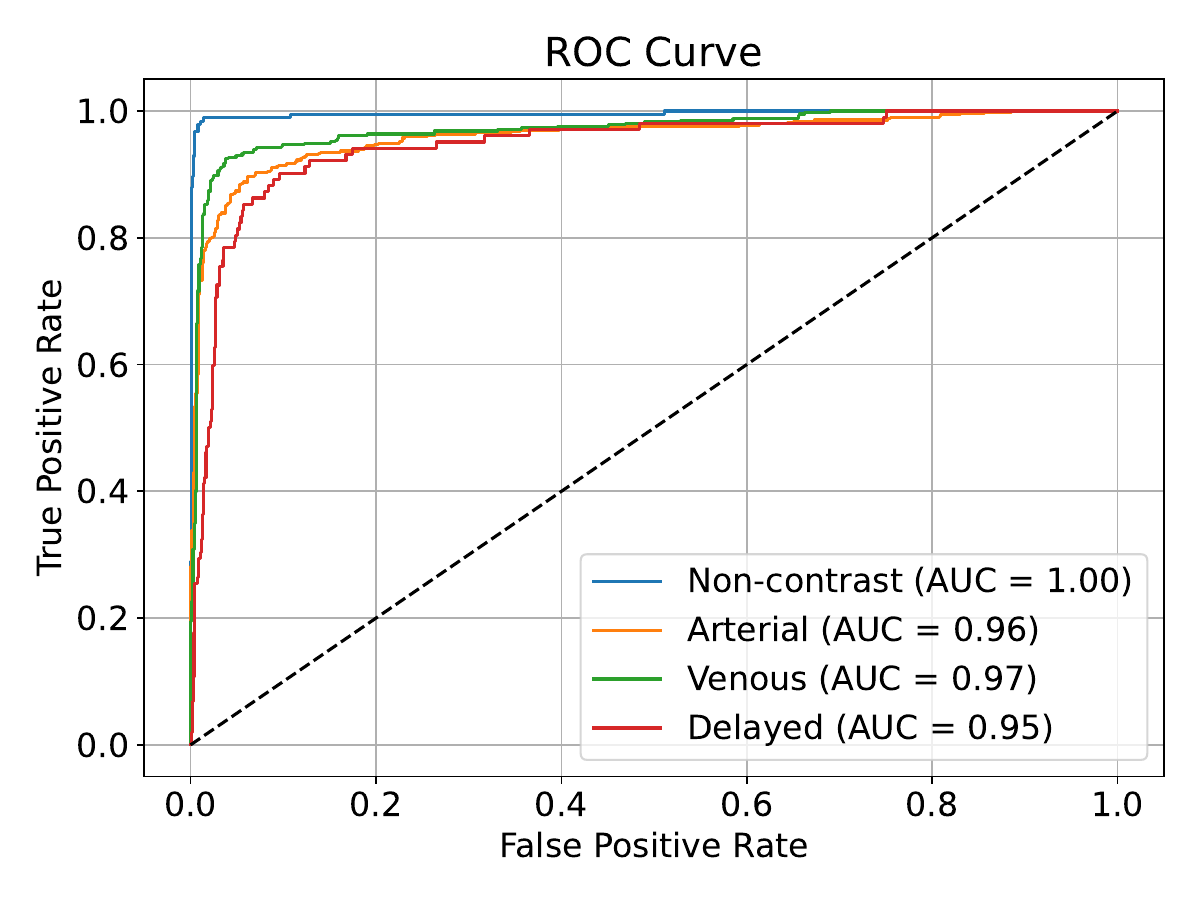} &
    \includegraphics[width=0.22\linewidth]{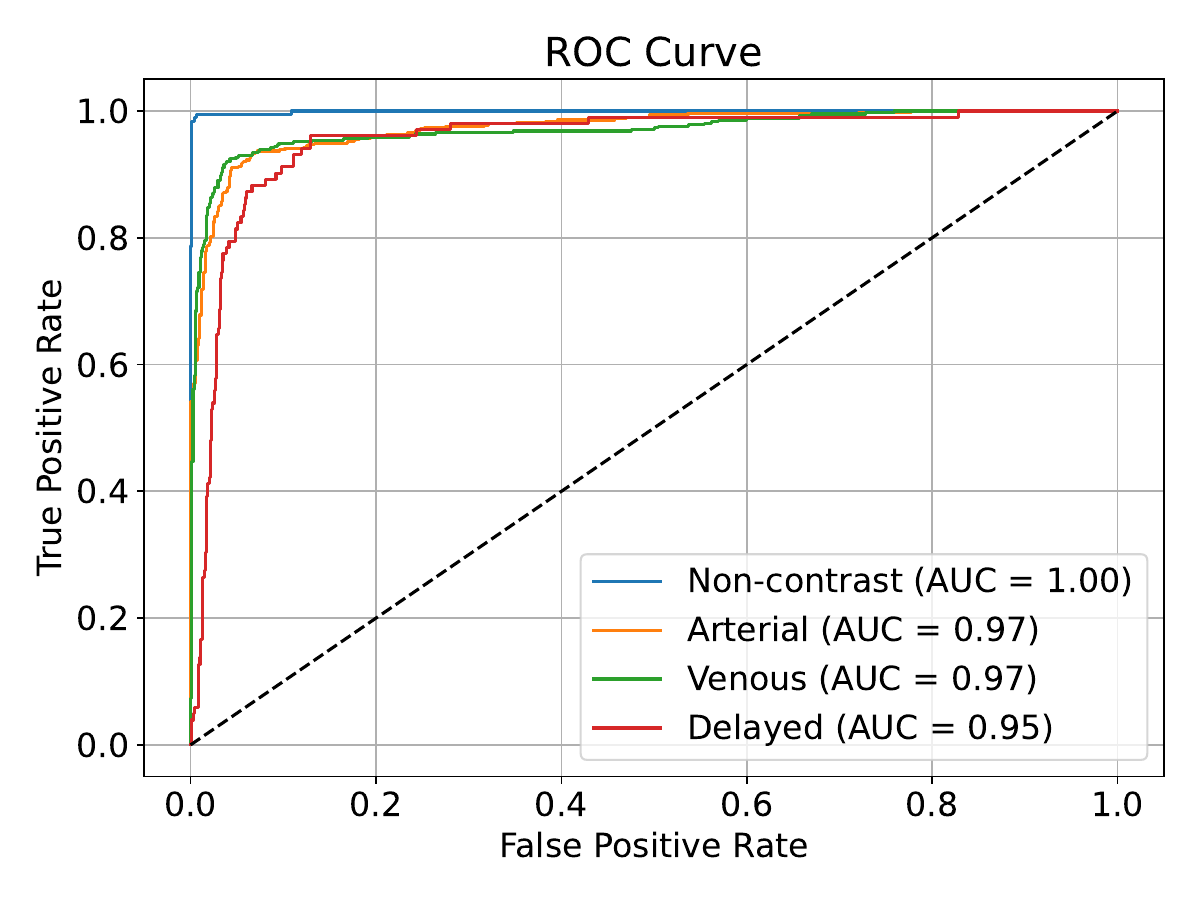} &
    \includegraphics[width=0.22\linewidth]{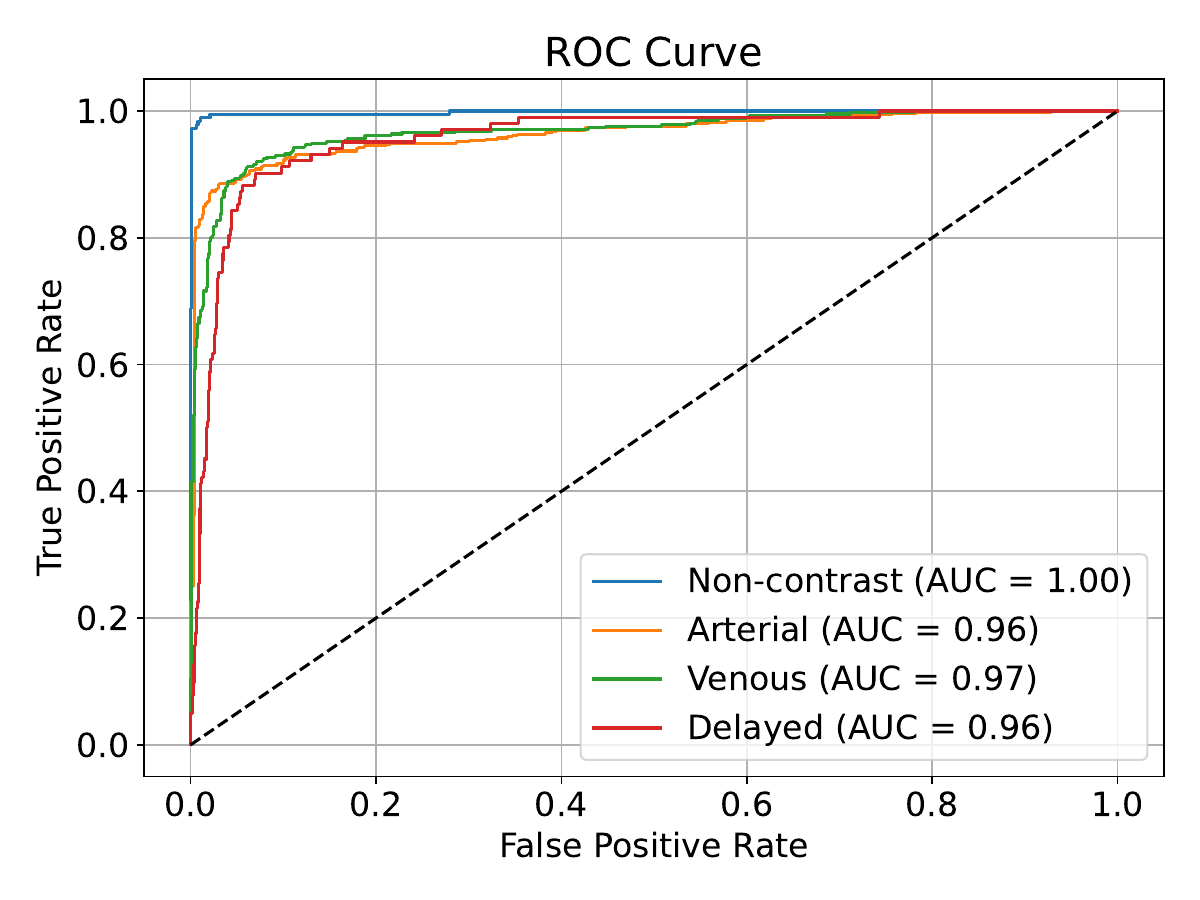} \\
    \small (a) XGBoost & \small (b) r3d\_18 & \small (c) mc3\_18 & \small (d) r2plus1d\_18
    \end{tabular}     
    \caption{Phase classification performance on the VinDr-Multiphase Dataset, illustrated by confusion matrices and ROC curves; (a) XGBoost, (b) ResNet3D 18-layer (r3d\_18), (c) Mixed Convolution Network 18-layer (mc3\_18), and (d) R(2+1)D 18-layer (r2plus1d\_18).}
    \label{fig:roc_cm_vindr}
\end{figure}

\begin{figure}[H]
    \centering
    \begin{tabular}{cccc}
    \includegraphics[width=0.22\linewidth]{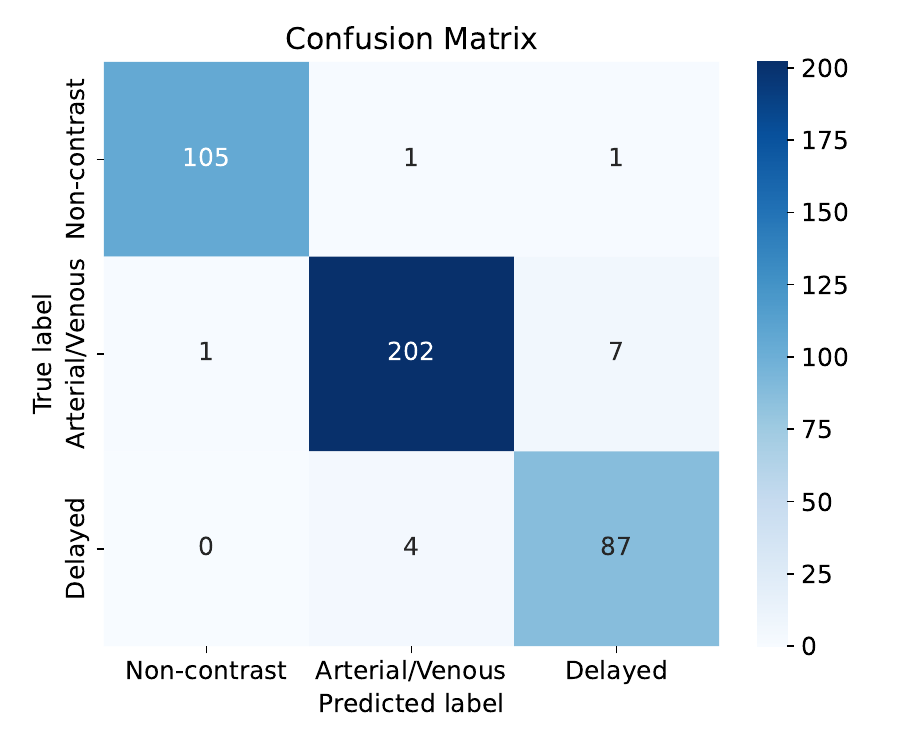} &
    \includegraphics[width=0.22\linewidth]{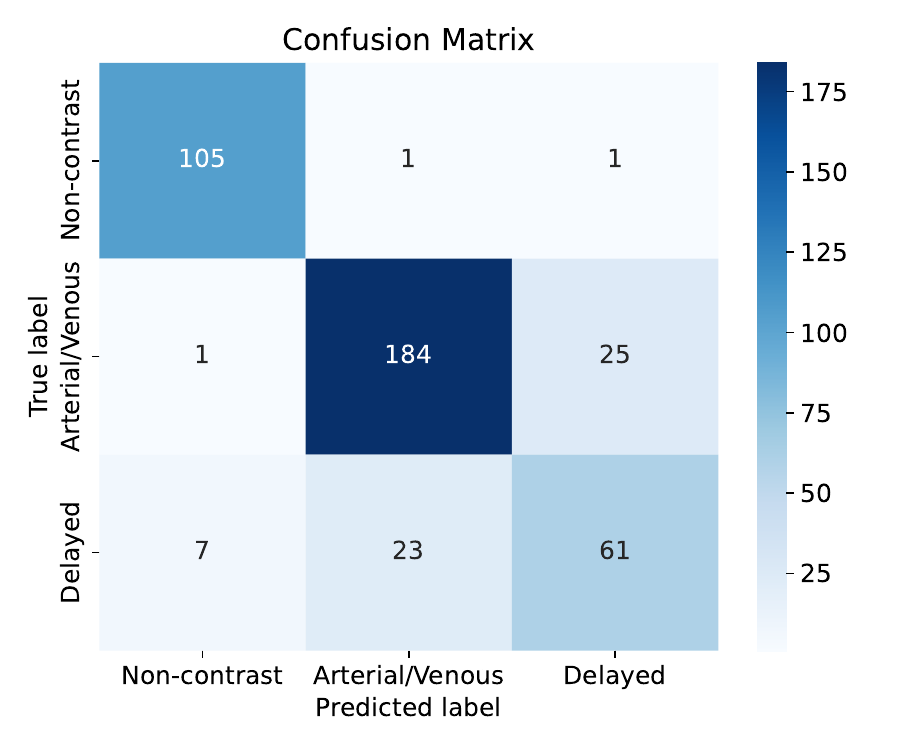} &
    \includegraphics[width=0.22\linewidth]{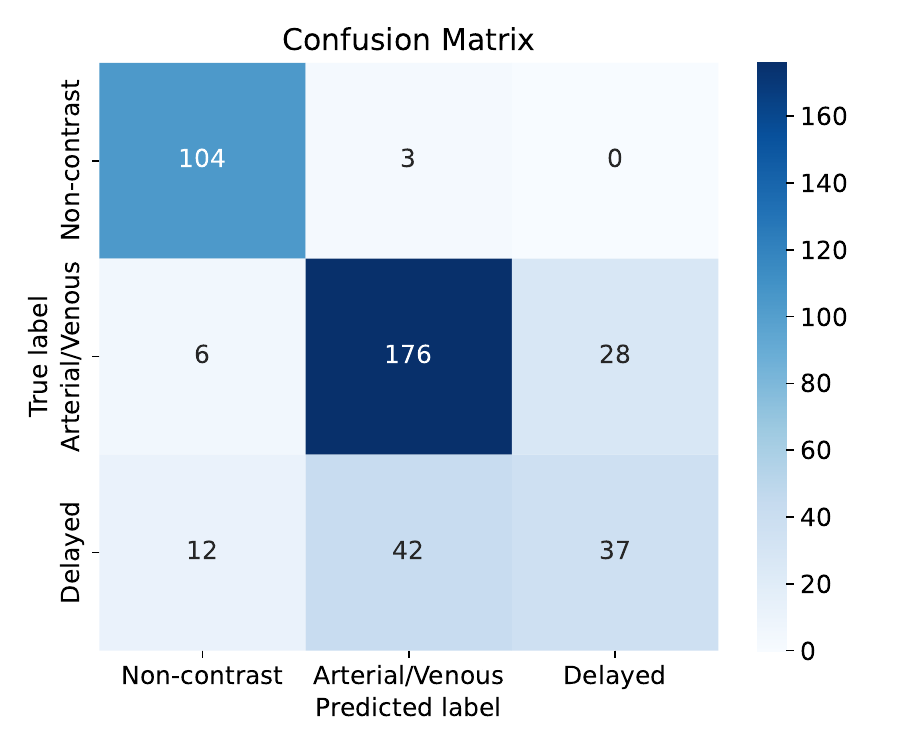} &
    \includegraphics[width=0.22\linewidth]{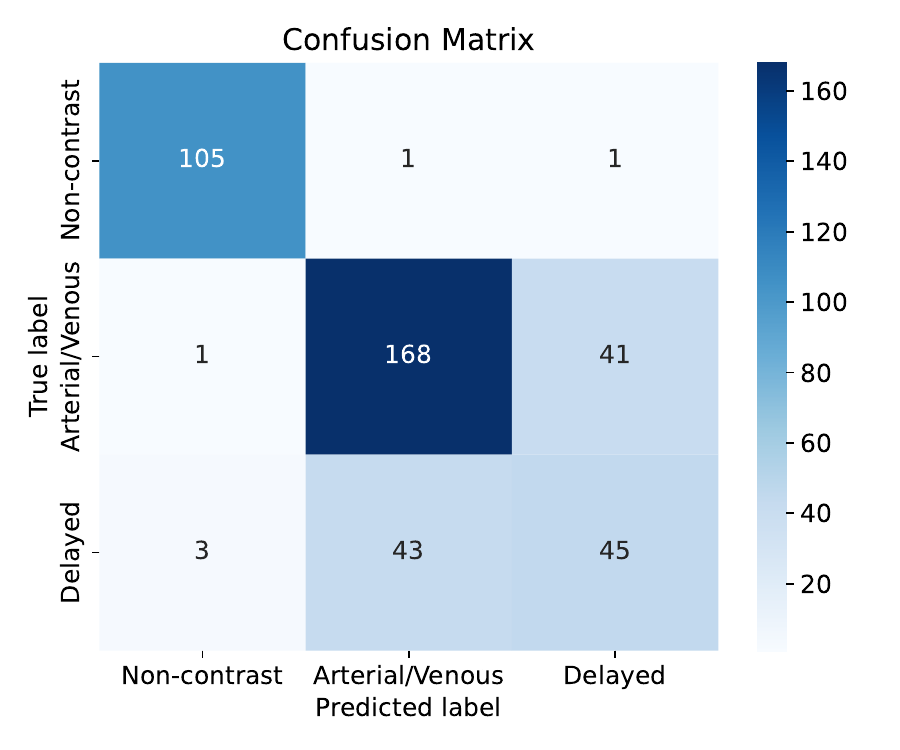} \\
    \includegraphics[width=0.22\linewidth]{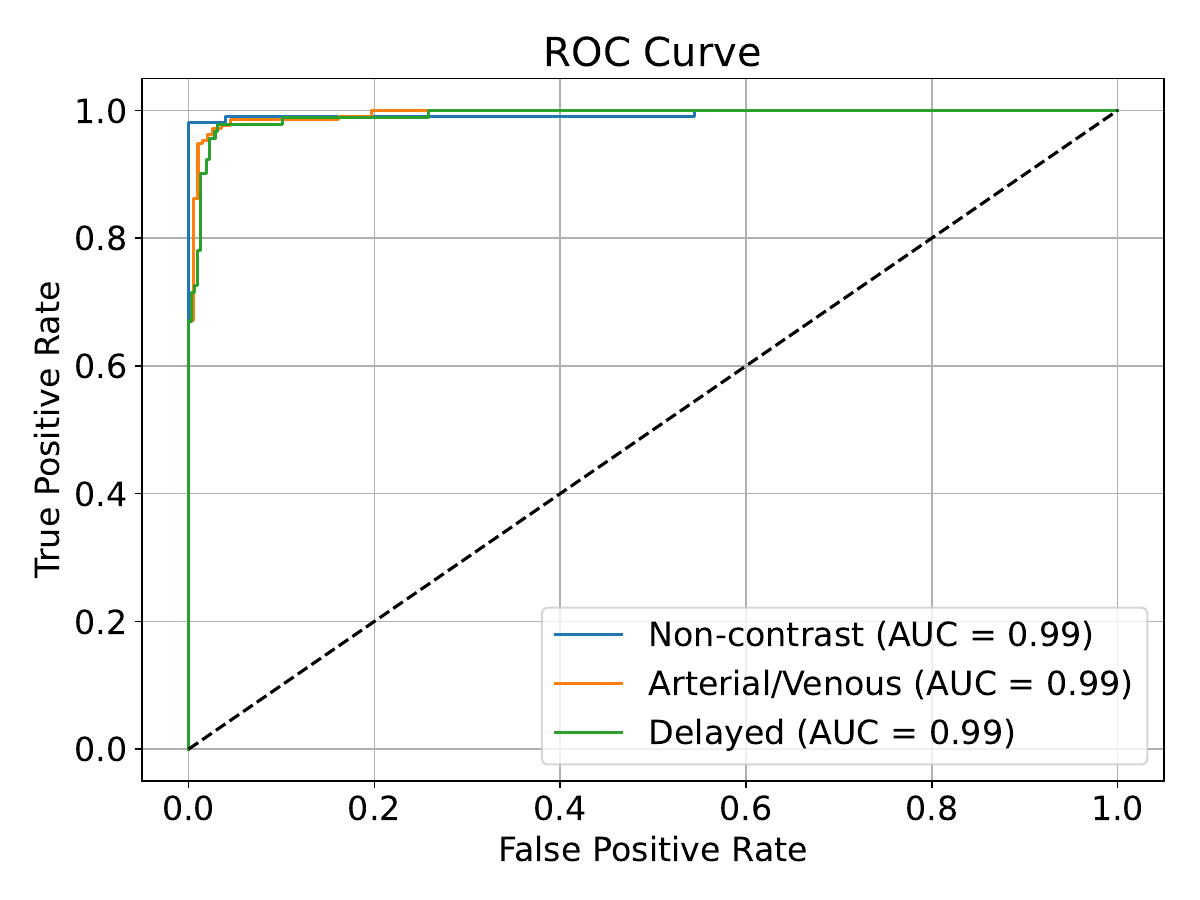} &
    \includegraphics[width=0.22\linewidth]{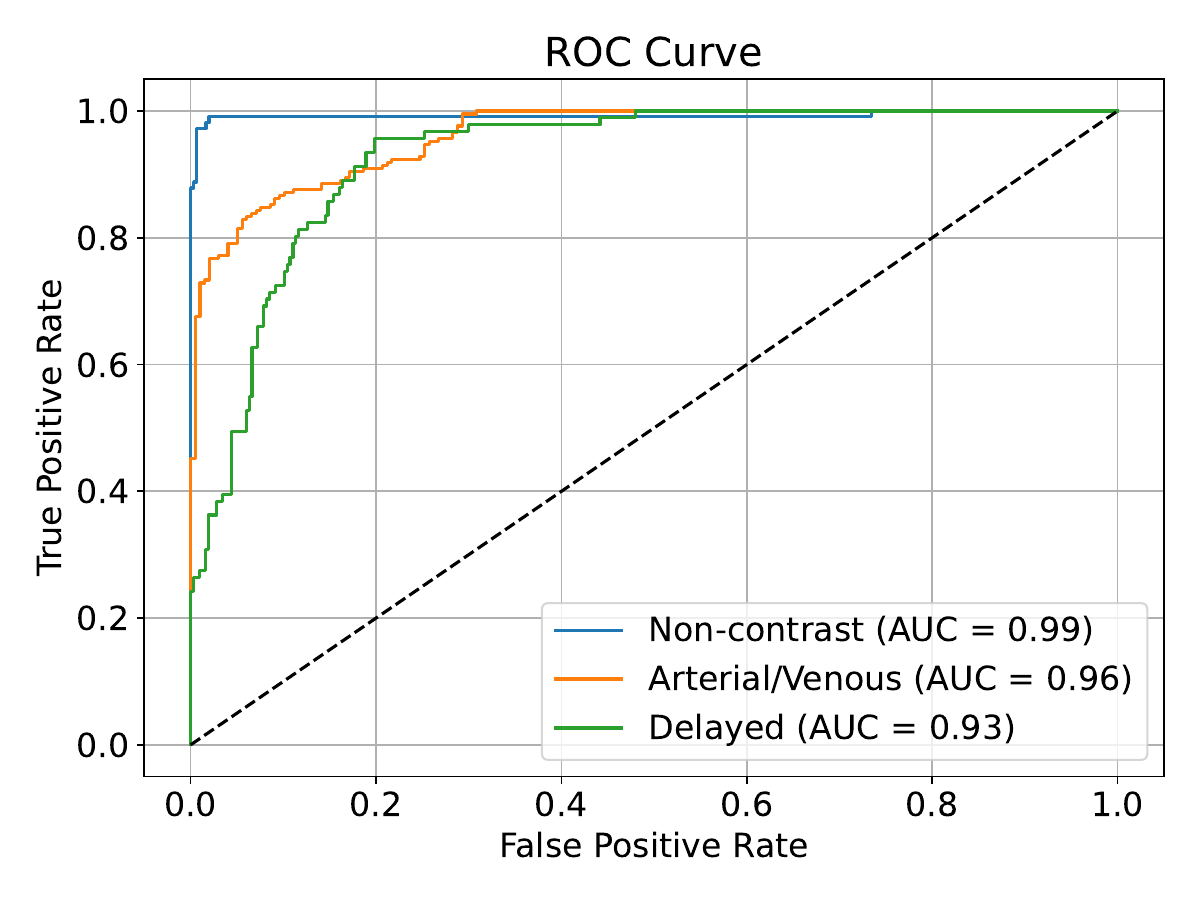} &
    \includegraphics[width=0.22\linewidth]{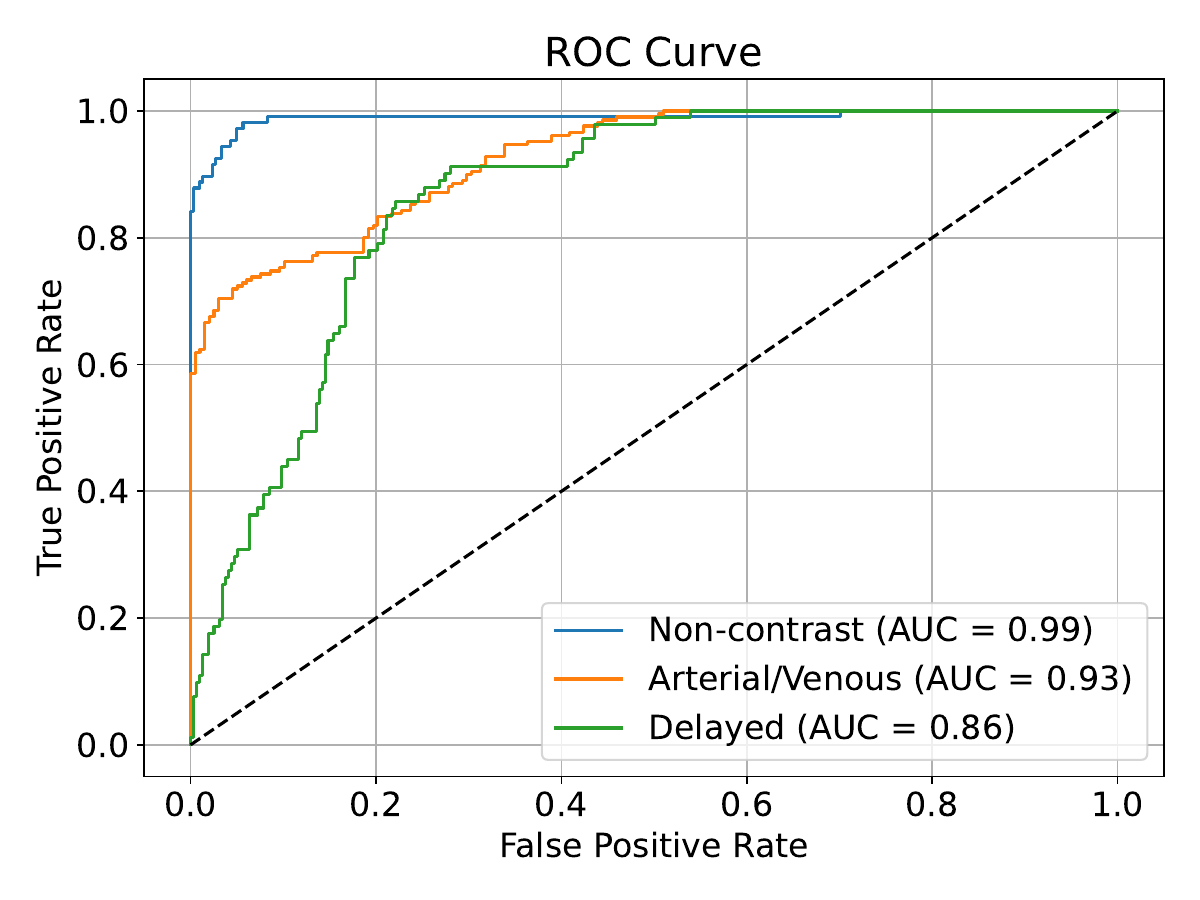} &
    \includegraphics[width=0.22\linewidth]{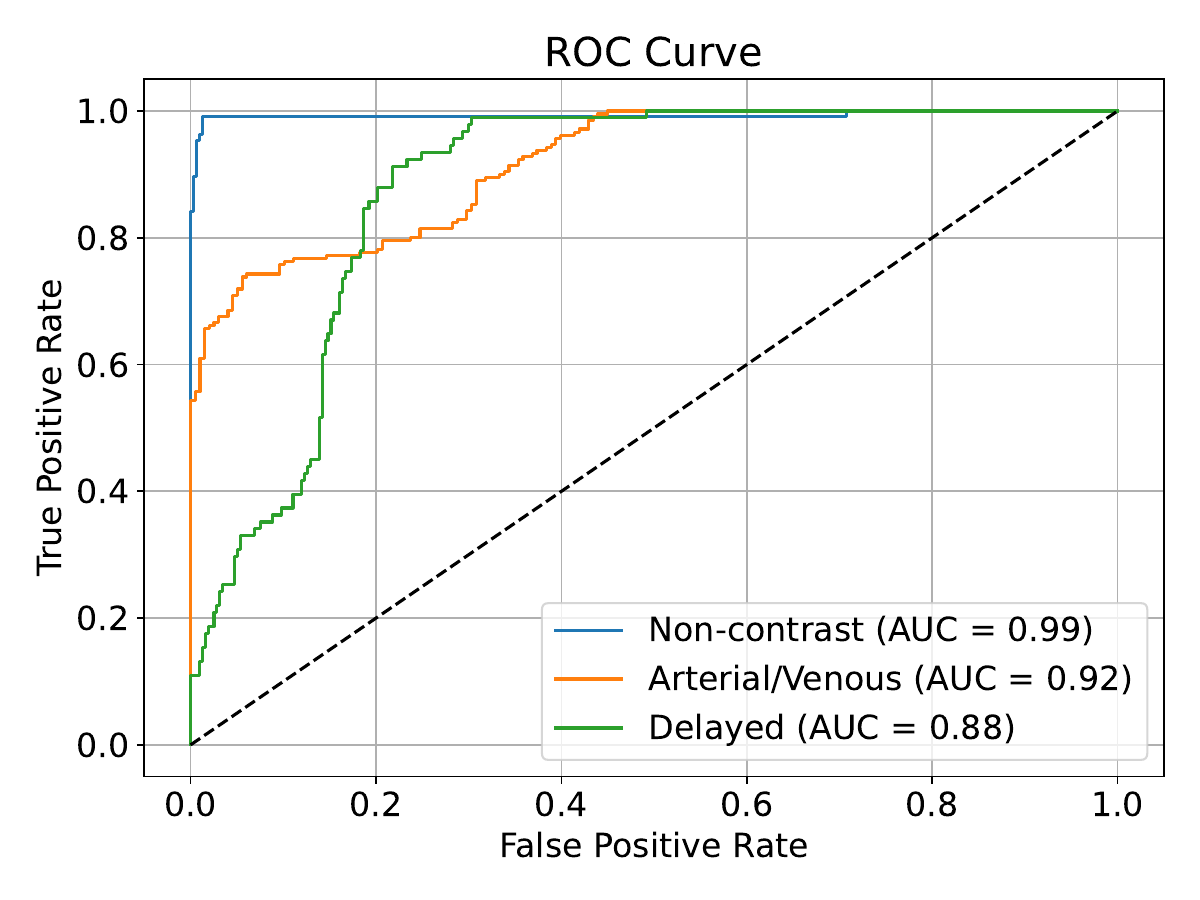} \\
    \small (a) XGBoost & \small (b) r3d\_18 & \small (c) mc3\_18 & \small (d) r2plus1d\_18
    \end{tabular}    
    \caption{Phase classification performance on the C4KC-KiTS Dataset, illustrated by confusion matrices and ROC curves; (a) XGBoost, (b) ResNet3D 18-layer (r3d\_18), (c) Mixed Convolution Network 18-layer (mc3\_18), and (d) R(2+1)D 18-layer (r2plus1d\_18).}
    \label{fig:roc_cm_c4kckits}
\end{figure}

\begin{figure}[H]
    \centering
    \begin{tabular}{cc}
    \includegraphics[width=0.22\linewidth]{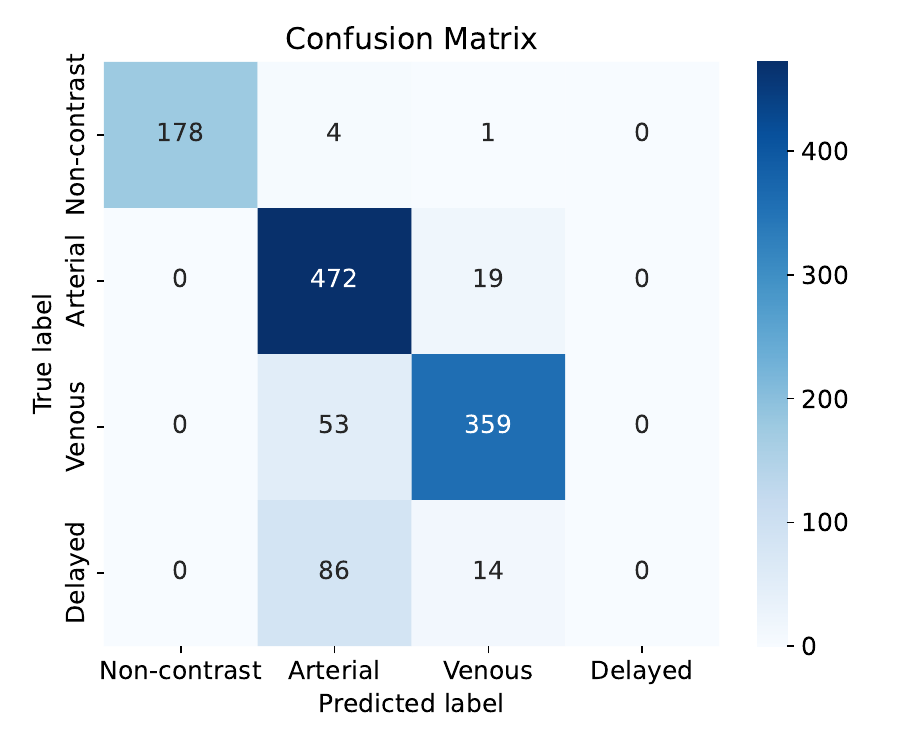} &
    \includegraphics[width=0.22\linewidth]{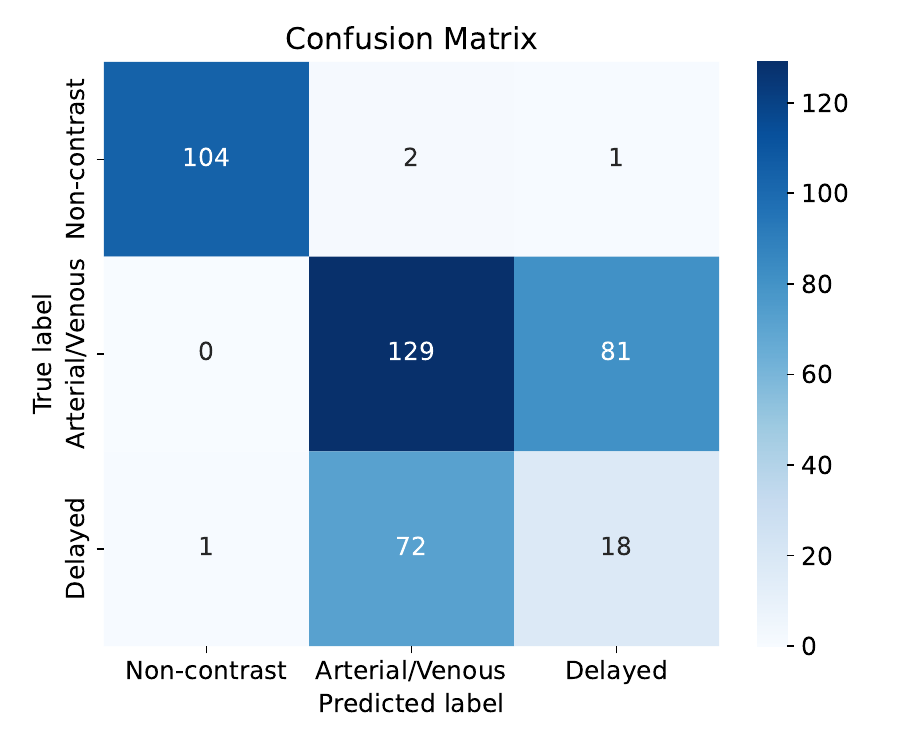} \\
    \small (a) VinDr-Multiphase & \small (b) C4KC-KiTS 
    \end{tabular}    
    \caption{Phase classification performance for ts\_phase on the (a) VinDr-Multiphase and (b) C4KC-KiTS Datasets.}
    \label{fig:cm_ts_phase}
\end{figure}

\end{document}